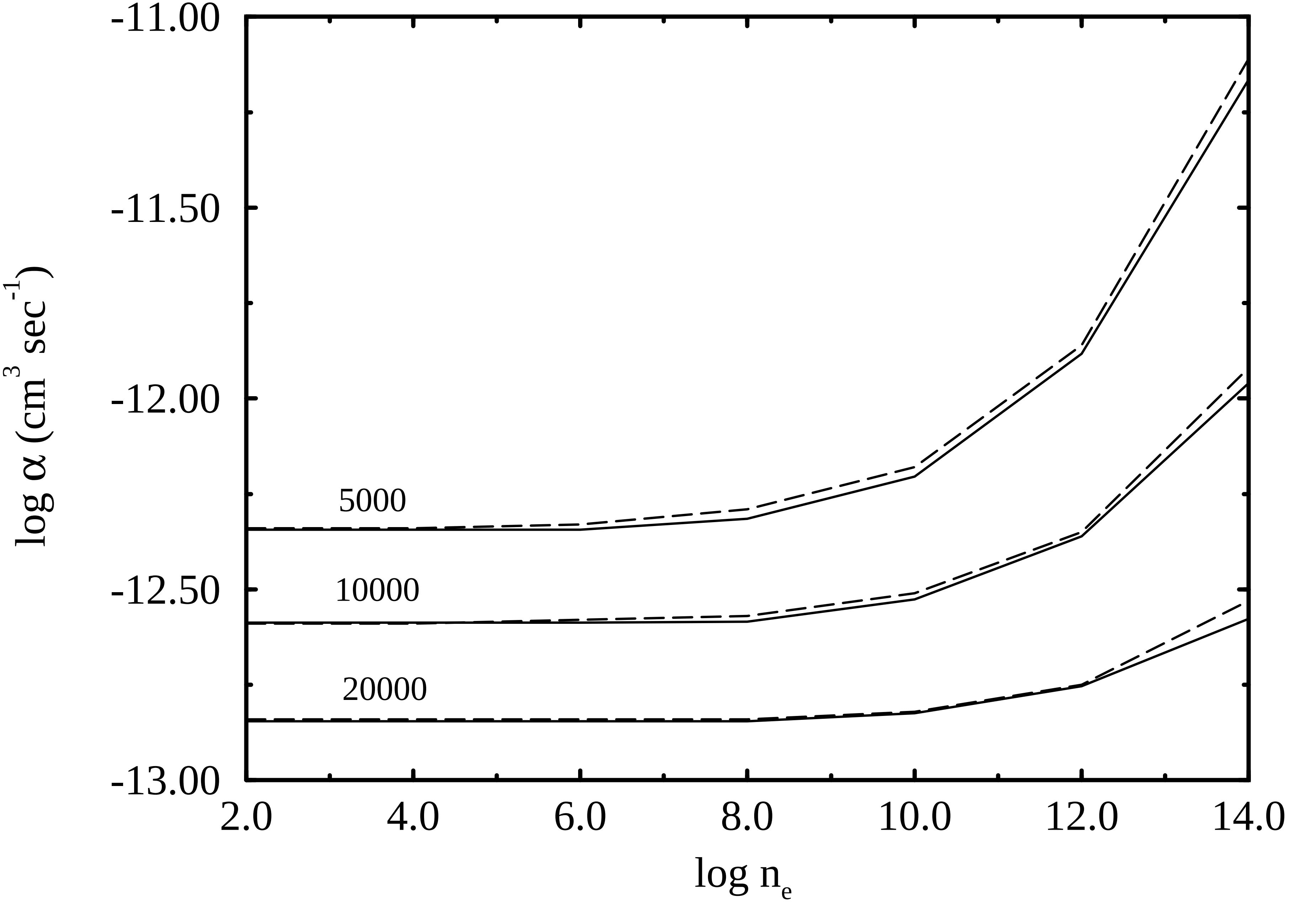

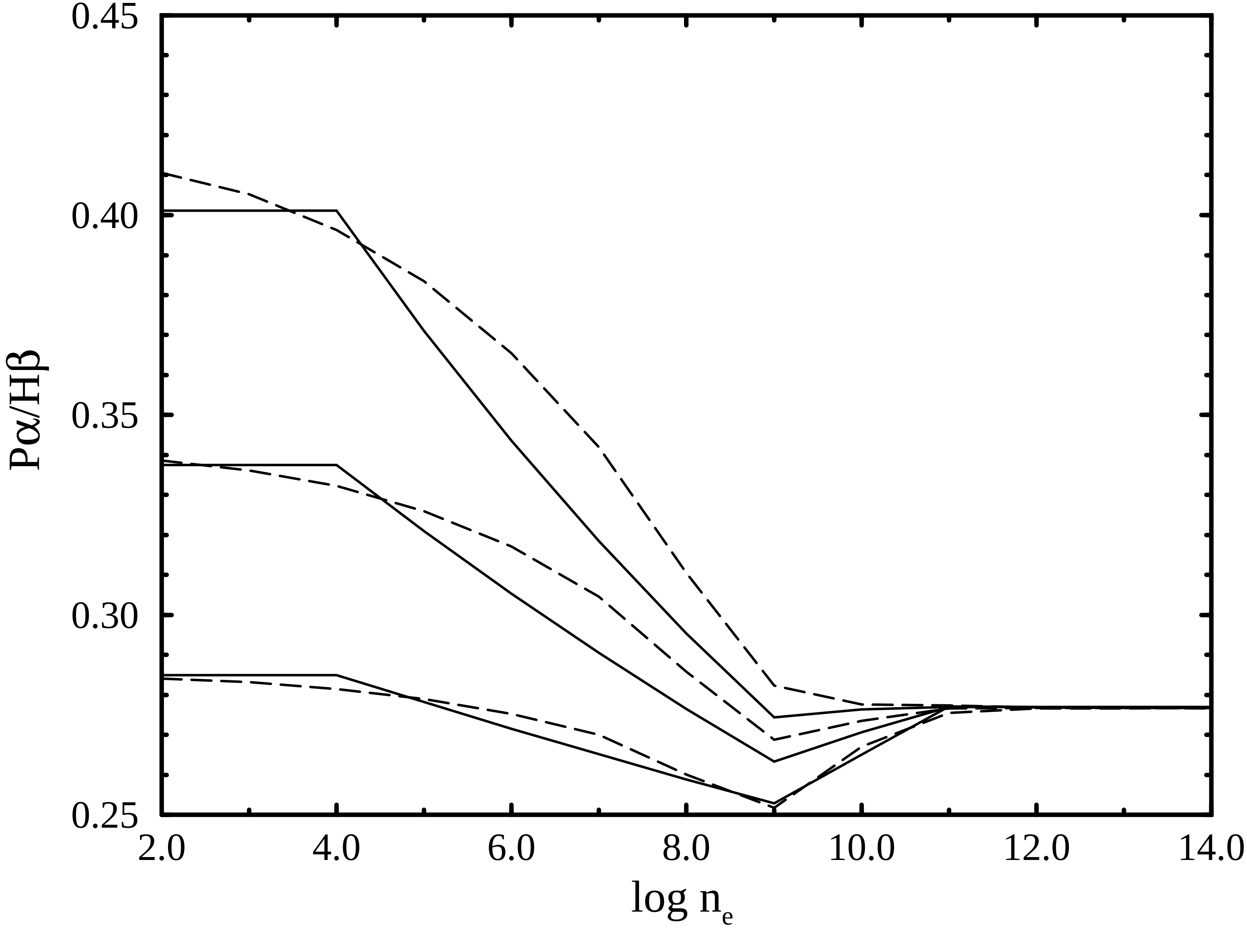

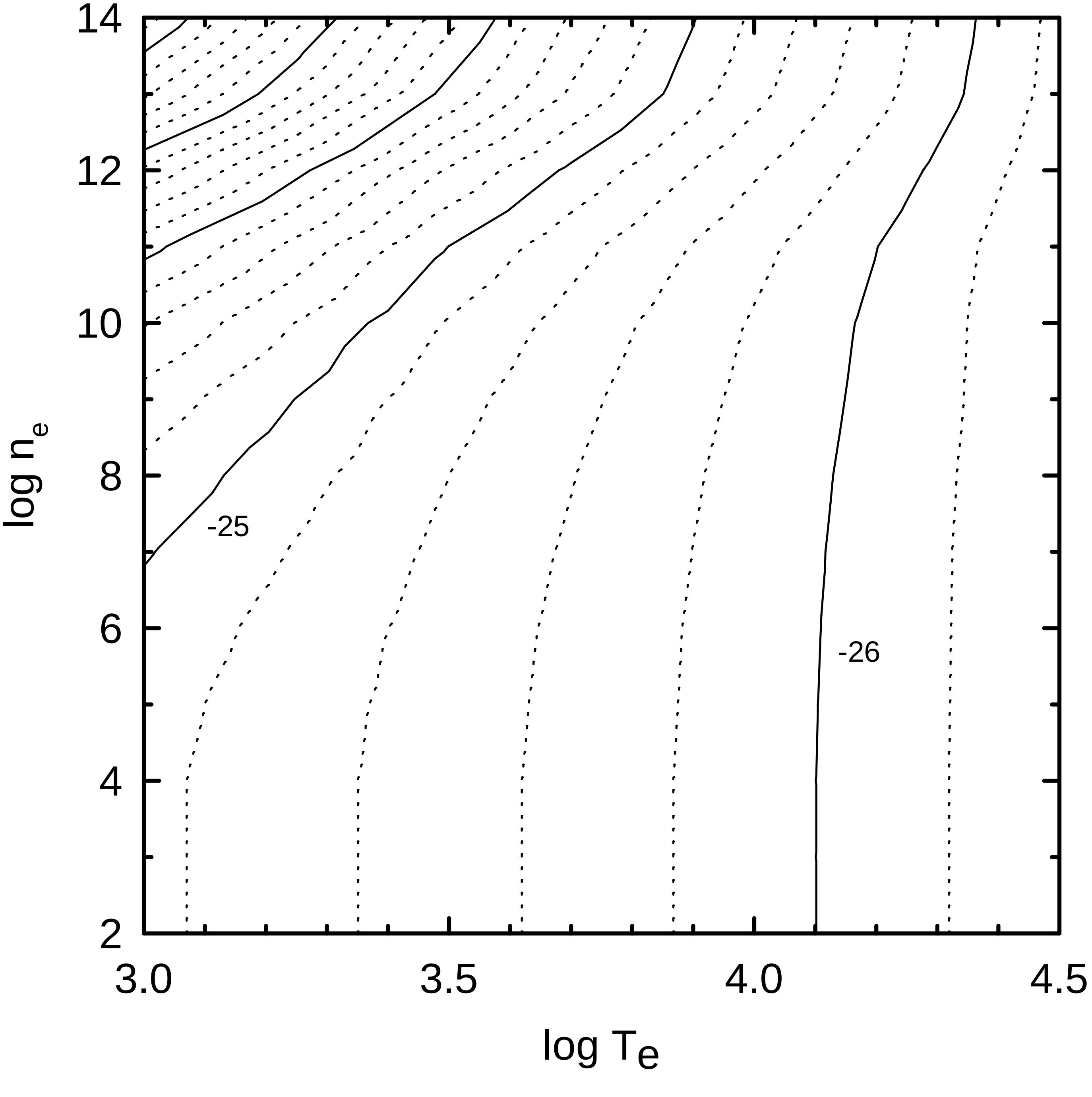

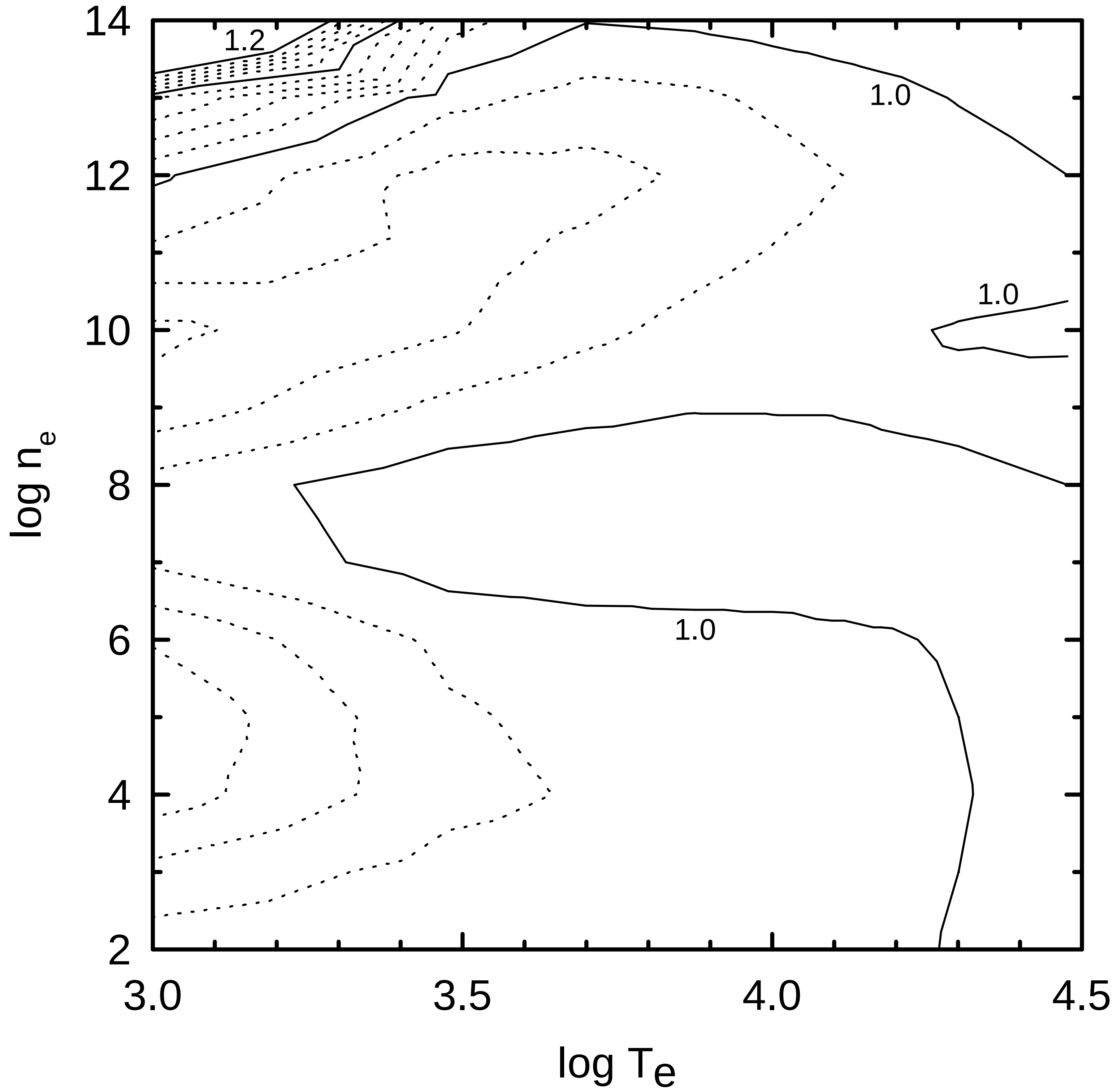

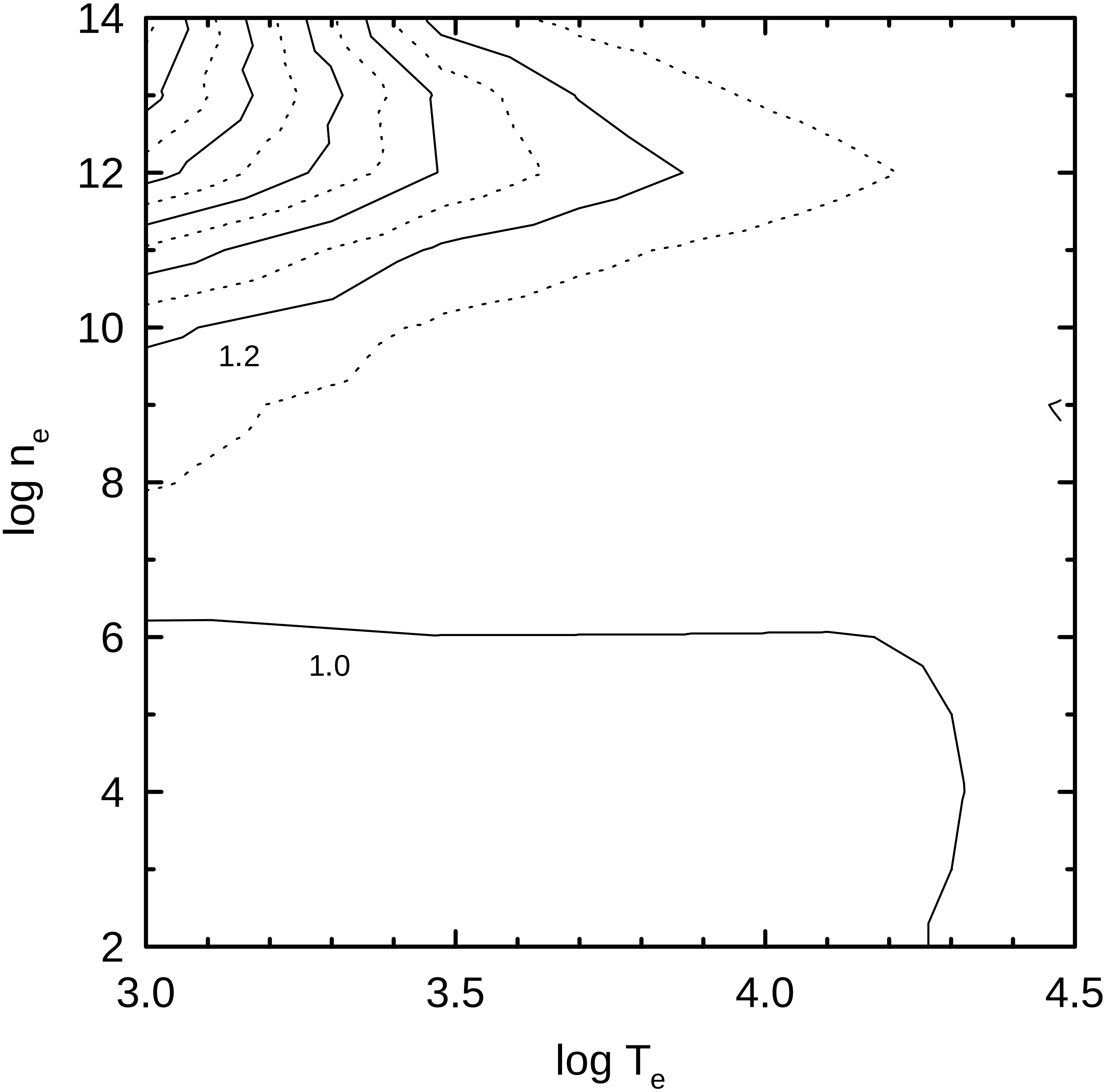

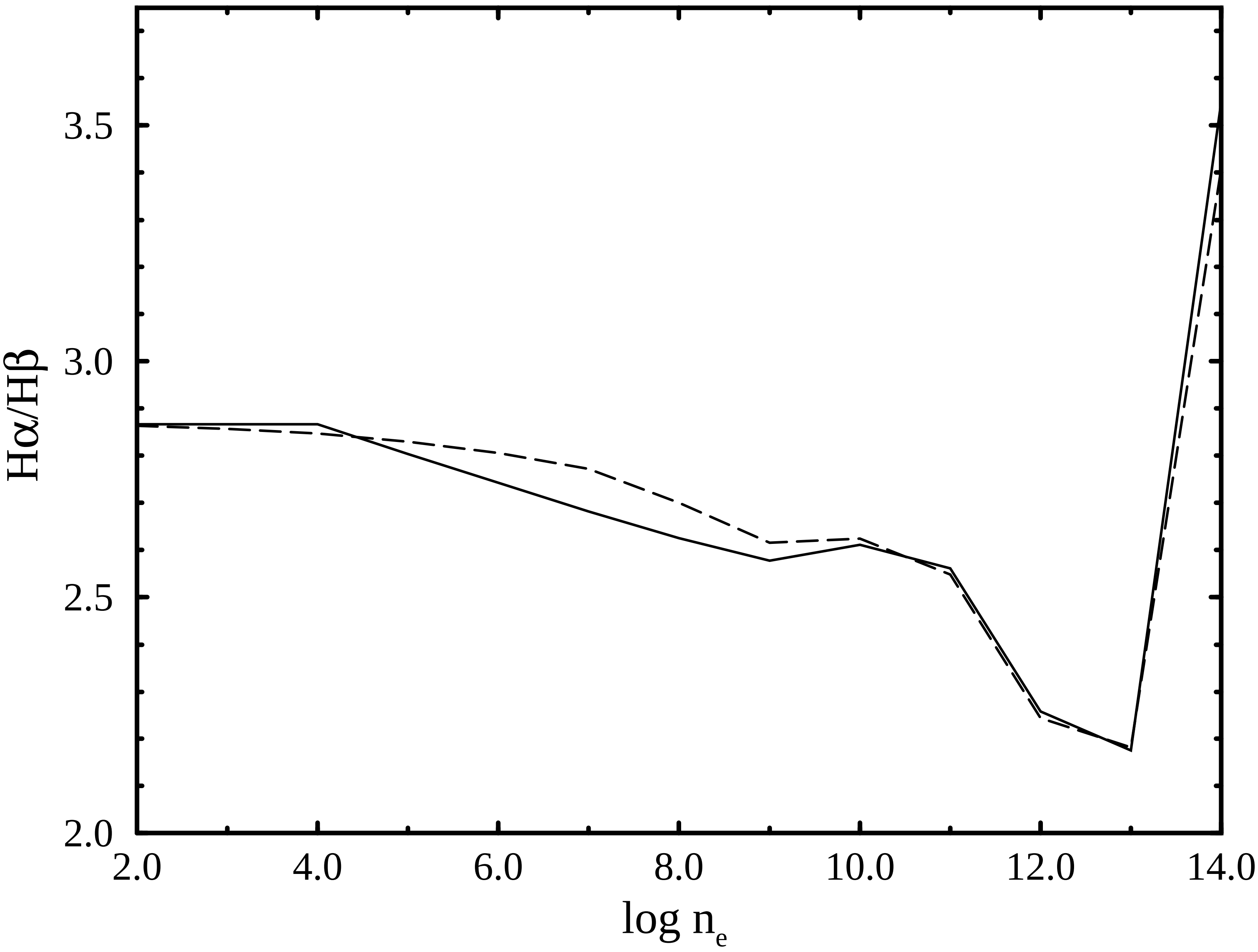

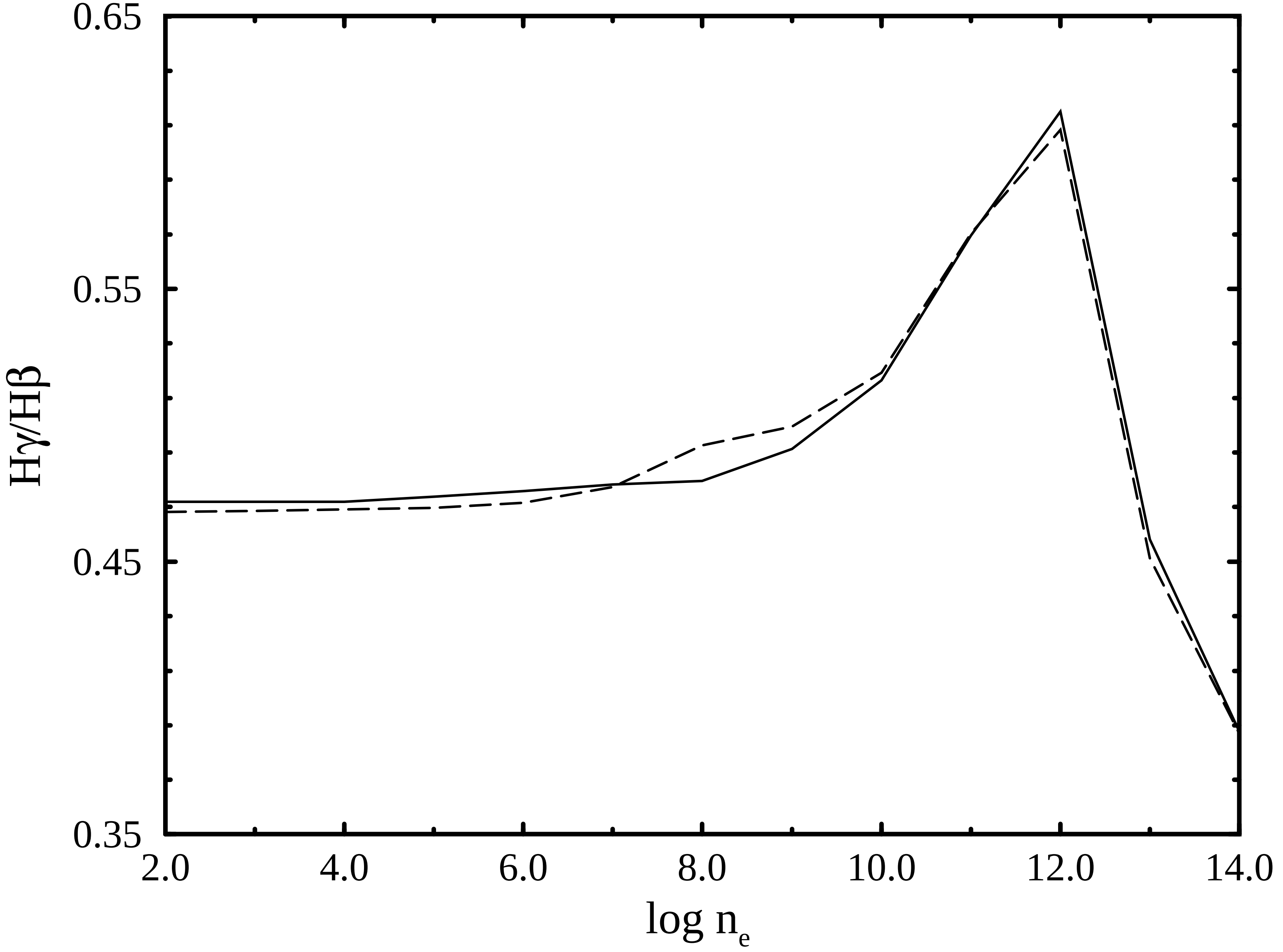

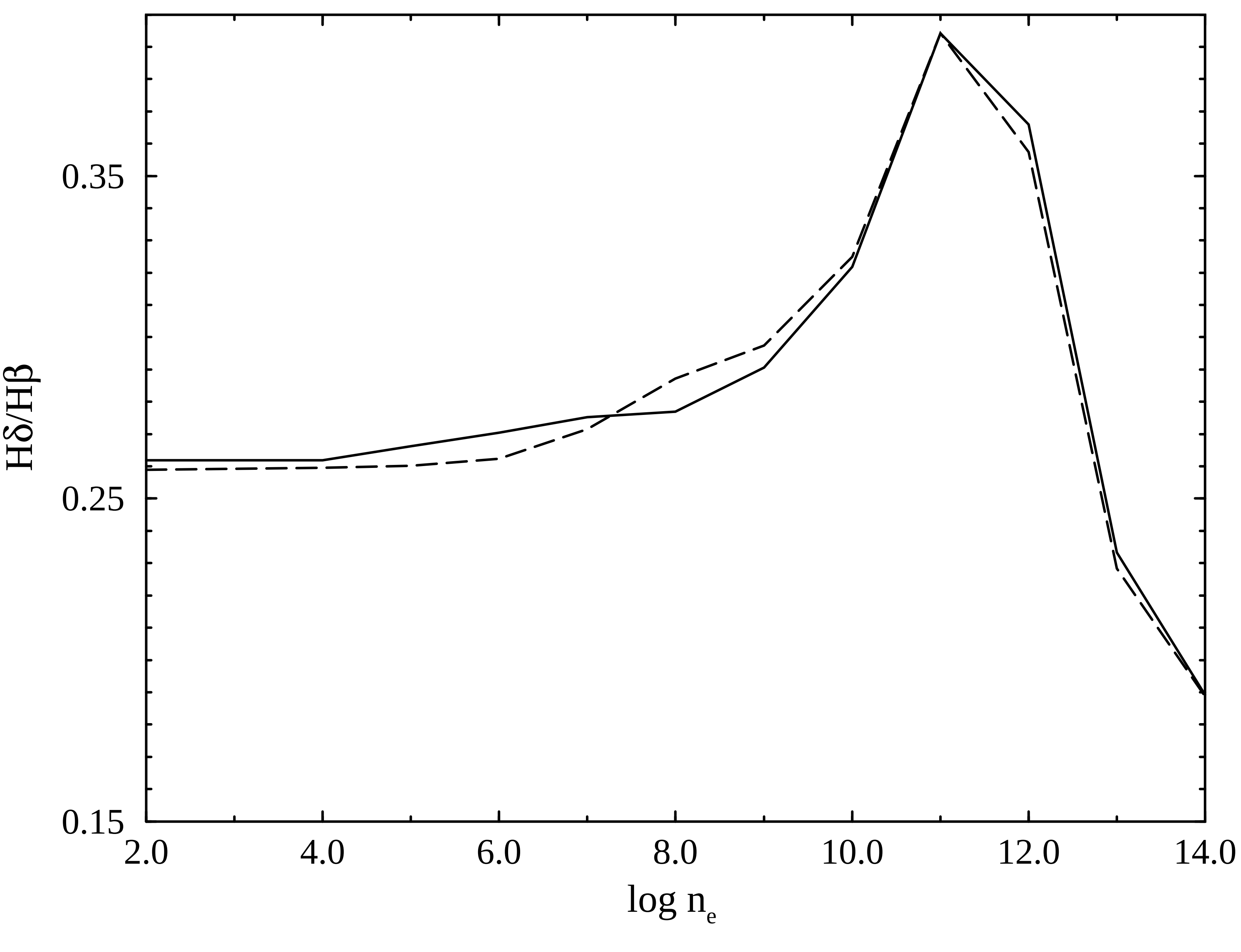

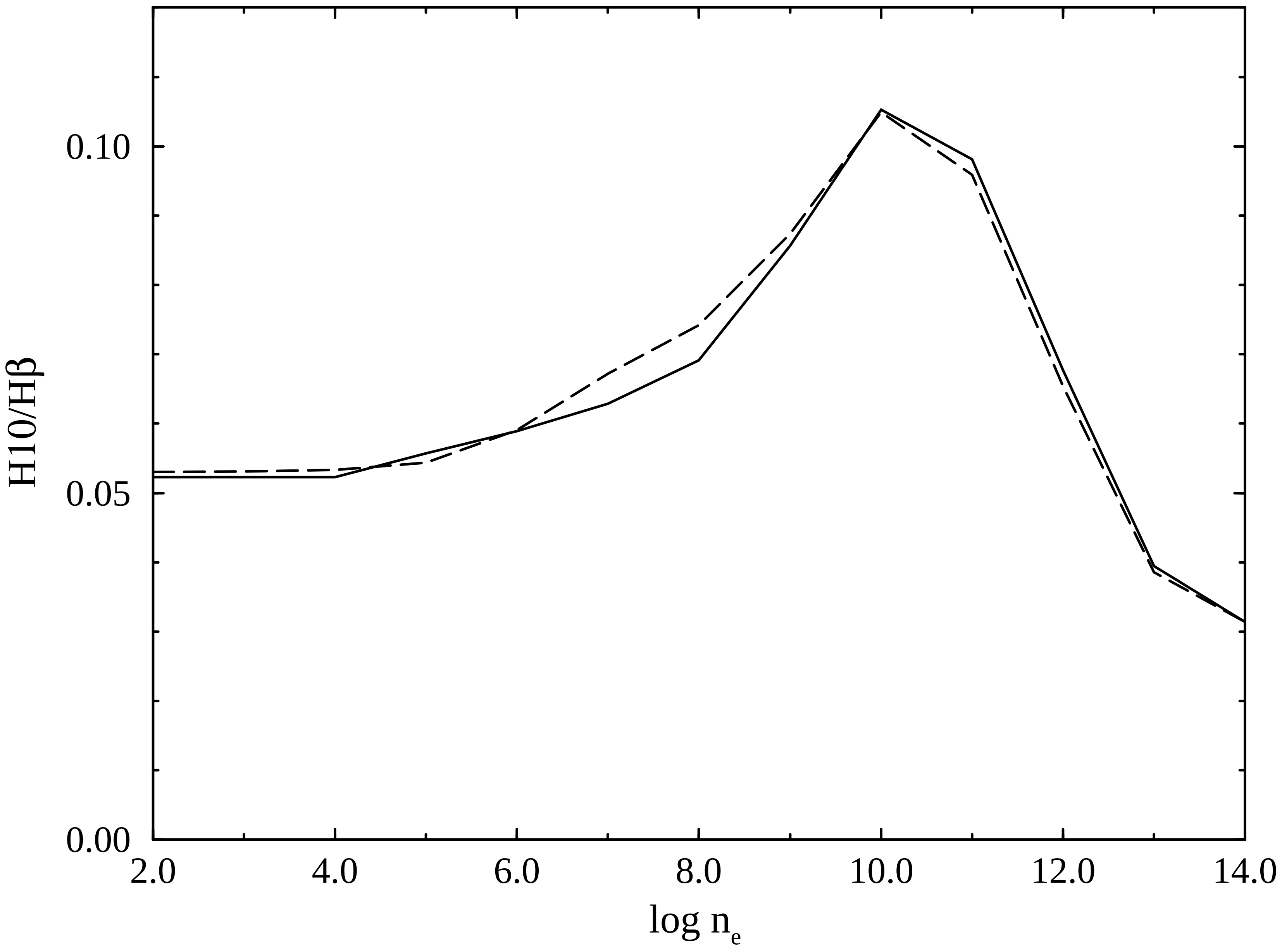

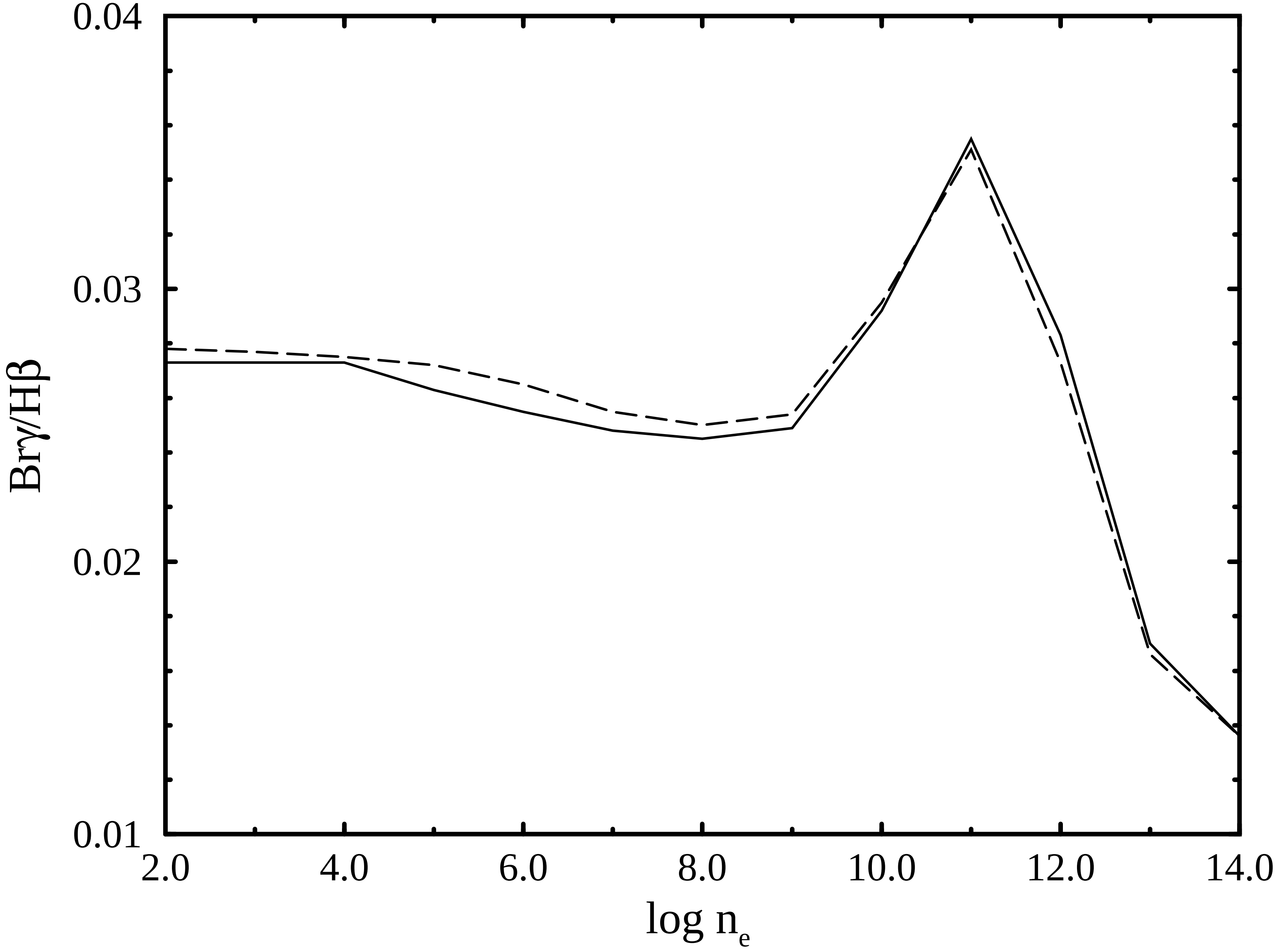

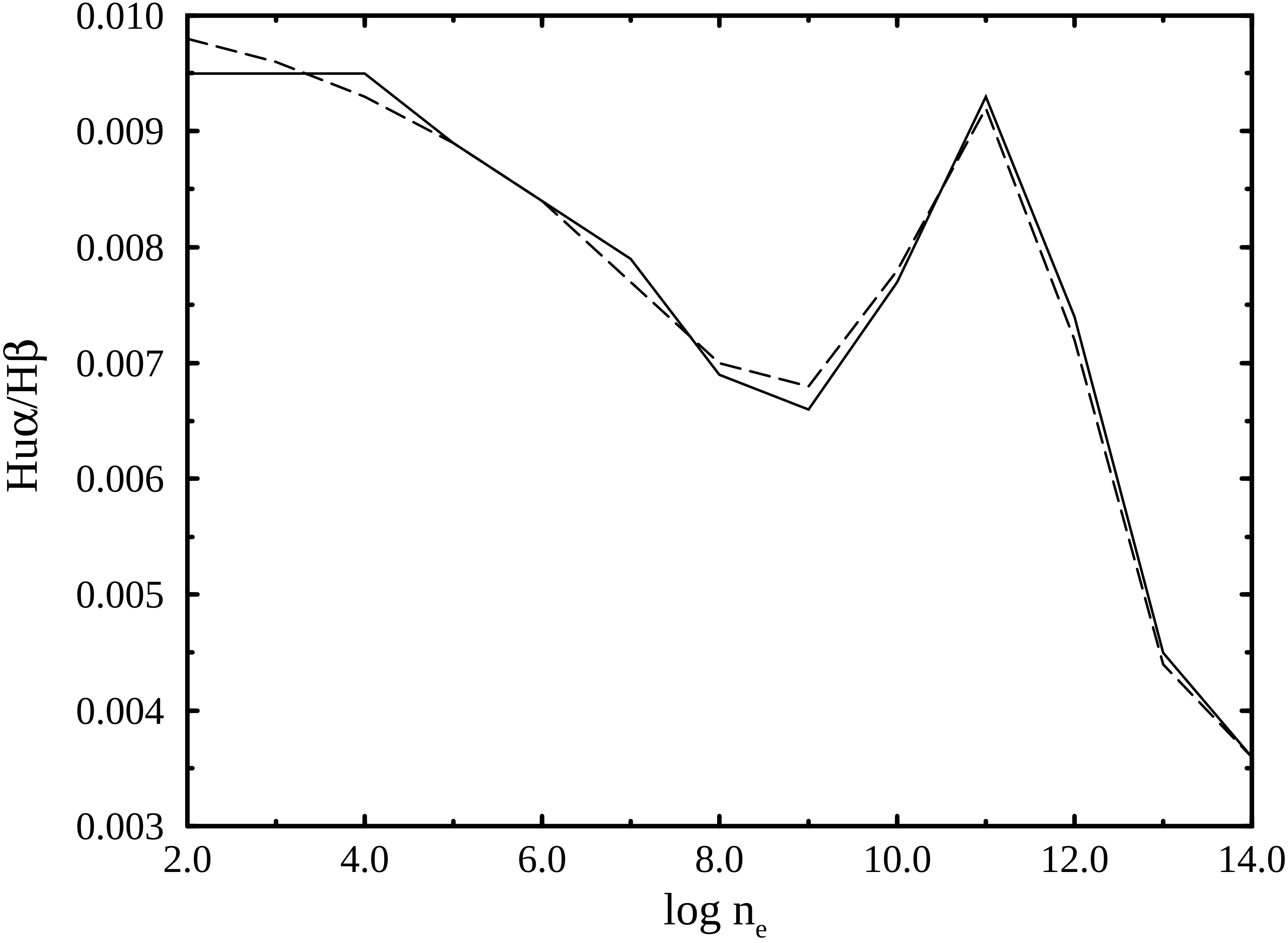

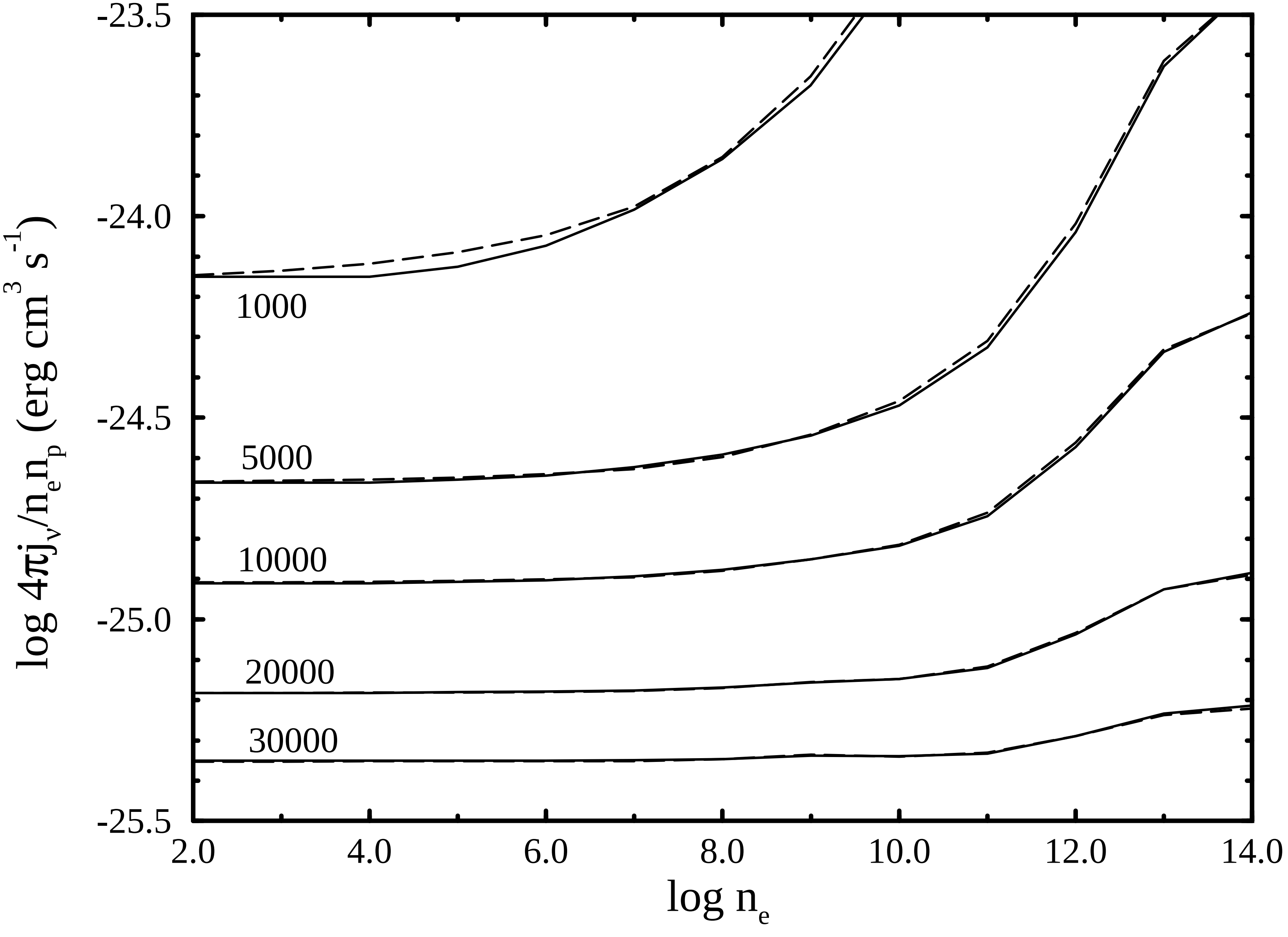

# Accurate Hydrogen Spectral Simulations with a Compact Model Atom

Jason W. Ferguson and Gary J. Ferland

Department of Physics and Astronomy

University of Kentucky

Lexington, KY 40506

Abstract

Many large scale numerical simulations of astrophysical plasmas must also reproduce the hydrogen ionization and the resulting emission spectrum, in some cases quite accurately. We describe a compact model hydrogen atom that can be readily incorporated into such simulations. It reproduces the recombination efficiency and line spectrum predicted by much larger calculations for a very broad range of densities and temperatures. Uncertainties in hydrogen collision data are the largest source of differences between our compact atom and predictions of more extensive calculations, and underscore the need for accurate atomic data.

## 1. Introduction

Hydrogen is the most abundant element and its physics dominates the conditions in a wide variety of galactic and extragalactic nebulae (Osterbrock 1989). Hydrogen's recombination efficiency and level populations must be computed with some precision since its photoionization can be the controlling heating mechanism for many clouds, it is the dominate opacity for many forms of light, and H is often the dominant electron donor (Avrett & Loeser 1988; Rees, Netzer, & Ferland 1989; Netzer 1990). The line spectrum must be predicted with great precision if reliable abundance determinations are to be achieved (Martin 1988; Shields 1990; Pagel 1992). Modern spectral synthesis codes (see Kallman & Mushotzky 1985; Binette et. al. 1993; Ferland et al. 1995, Netzer 1996) must incorporate complete simulations of the H atom if they are to be valid.

A number of physical processes affects hydrogen, including radiative and three-body recombination, photoionization and induced recombination, and collisional and radiative (spontaneous and induced) transitions between levels. Complete calculations of the physics of the hydrogen atom are intricate themselves (Brocklehust 1971; Mathews, Blumenthal & Grandi 1980, Drake & Ulrich 1980; Storey & Hummer 1995), and certainly cannot be incorporated into larger structure and spectral synthesis codes. These processes are important in the Broad Line Region (BLR) of Quasars or any dense gas, and so must be fully simulated.

Storey & Hummer (1995 hereafter SH), have completed full calculations of the hydrogen emission spectrum and recombination efficiency over a very broad range of temperature and density. Their calculations are for "Case A" and "Case B" conditions, with the 1000 lowest $n$ levels and all $l$ states included. These computations are likely to remain definitive for some time to come, and they made their results available electronically, along with a convenient interpolating program. Unfortunately, it is possible to incorporate model hydrogen atoms with only dozens, not hundreds or thousands, of levels, into full spectra synthesis simulations. The problem is to define a compact model atom which retains the essential physics and agrees with the definitive calculations.

This paper outlines a model atom that can accurately reproduce the recombination efficiency and line emission with only a few dozen quantum levels, and can be incorporated into large-scale structure simulations. The approach goes to the correct asymptotic limits, and reproduces accurate results in intermediate conditions. We pay particular attention to obtaining accurate representations of emissivities of infrared lines. Surprisingly, the rates for collisions within the hydrogen atom appears to be the limiting factor on accuracy. Finally, we compare the idealized Case B emissivity with results from a realistic BLR cloud model.

## 2. Physical Processes, and their simulation with a compact atom

Our model hydrogen atom consists of independent 1s, 2s, and 2p states, and an arbitrary number of higher excited n-levels.

Both the low-density and LTE limits, as well as the intermediate "Case A" and "Case B" are well defined in the literature. Case B comes closest to simulating nature, and we concentrate on it here. For this approximation, all transitions involving the ground state are ignored, as is photoionization from excited states, collisional ionization and excitation from the ground and first excited states, fluorescent excitation, and all induced processes. These are normally included in our calculations but are disabled in the work presented below.

### *2.1. Photoionization, radiative recombination*

Recombination to an infinite sum of bound levels must be included if the total recombination coefficient, and resulting ionization balance, are to be correct.

We have fit radiative recombination coefficients for levels up to 15 using the Milne relation and photoionization cross sections taken from Opacity Project-related codes, as described in Ferland et al. (1992). The fit has the functional form

$$\alpha(n,T) = 10^{F(n,T)}\ T_e^{-1} \ ,$$

where

$$F(n,T_e) = \frac{a_n + c_n x + e_n x^2 + g_n x^3 + i_n x^4}{1 + b_n x + d_n x^2 + f_n x^3 + h_n x^4} \ , \tag{1.}$$

and $x$ = log $T_e$. Table 1 gives the coefficients used for the above equation, which are valid for the temperature range 2.8 K ≤ $T_e$ ≤ $10^{10}$ K.

For quantum levels above 15 we use an asymptotic analytical formula from Allen (1973):

$$\alpha_n = \left(3.262\times10^{-6}\right) n^{-3}\ T_e^{-3/2} \exp\left(\frac{\chi_n}{kT_e}\right) E_1\left(\frac{\chi_n}{kT_e}\right), \tag{2.}$$

where $\chi_n$ is the ionization potential of the nth level and $E_1$ is the first exponential integral (Abramowitz and Stegun 1965).

Errors in individual recombination coefficient fits are generally 1-2% with many less than 1% for lower excited states at temperatures of 10,000 K. For levels 1 ≤ $n$ ≤ 15 the largest errors are 5% at only the highest temperatures. Equation 2, used for levels $n$ ≥ 16, breaks down at temperatures greater than $10^8$ K, however at these high temperatures less that 2% of the total recombinations are to levels greater than 16, so this does not introduce an error in the total recombination coefficient at these high temperatures.

The approach given above only provides rates for levels explicitly included in our model atom, and thus is an underestimate of the total recombination coefficient, a sum to infinity. The total recombination coefficient must be accurately computed to produce accurate ionization representations in the low density limit. To compensate the highest 5 levels of the atom are assigned the remainder of the recombination coefficient:

$$\alpha_N = \alpha_B - \sum_{n=2}^{n=N-1} \alpha_n \ , \tag{3.}$$

where $\alpha_B$ is the total Case B rate, (the sum of recombinations to *n*=2 and higher levels fit with Equation 1 and coefficients listed in Table 1), *N* is the total number of levels in the compact model atom, and $\alpha_n$ is the recombination coefficient to the n$^{th}$ level. Although assigning the remaining recombination coefficient to the higher levels had the desired effect of including all the possible recombinations, it also resulted in the overpopulation of these levels. This caused a level inversion and resulted in a strong maser, and so was unphysical. We topped off the atom with a band of levels rather than the highest level to minimize this effect. The result of this topping off of the atom is that the model atom reproduces the $\alpha_B$ sum and the ionization balance at low densities to very great precision.

### 2.2. Collisional ionization, three body recombination

Collisional ionization and its reverse process, three body recombination, are important since they bring the atom into LTE at high densities. Collisional ionization from the ground state is taken from Arnaud & Rothenflug (1985). For levels 2 and higher we use rate coefficients of Vriens & Smeets (1980) who give a semi-empirical fit between theory and experiment for hydrogenic atoms. If the atom is in an initial excited state $|q\rangle$ with energy $E_q = 13.6\,eV/q^2$ then the rate coefficient for ionization is

$$K_q = \frac{9.56\times10^{-6}\left(kT_e\right)^{-1.5}\exp(-\varepsilon_q)}{\varepsilon_q^{2.33} + 4.38\,\varepsilon_q^{1.72} + 1.32\,\varepsilon_q}, \qquad (4.)$$

where $\varepsilon_q = E_q/kT_e$ is the Boltzmann factor with $kT_e$ in eV, and the equation is good for the values of $kT_e$ given by Vriens & Smeets (generally for temperatures greater than $10^3$ K). For reference, the definitive SH calculations use collision data from Burgess & Percival (1968). Chang, Avrett, & Loeser (1991) discuss the uncertainties of the collision data for hydrogen. For intermediate levels this is easily a factor of two.

Figure 1 shows the total recombination coefficient for three temperatures and a wide range of density. The solid line presents the results with a 50 level atom, and the dashed line SH. At low densities the recombination coefficient is purely radiative, while three-body recombination dominates the total at high density. For low temperatures and intermediate densities the differences can be as great as 5%. Tests discussed below show that the differences in our total recombination coefficient and those of SH are due to the different collision data assumed, and so are a basic uncertainty.

### 2.3. n-changing collisions

For collisional de-excitation involving the ground, first and second excited states of hydrogen we use data from Callaway (1994); excitation rates are calculated by the principle of detailed balance. Collisions between 2s and 2p are taken from Osterbrock (1989). For collisional de-excitation involving levels 4 and higher we again use the formulae of Vriens & Smeets (1980) which is available in closed form, is recommended by Chang, Avrett, & Loeser (1991) , and can be extended to any transition from any excited state. The atom is in an initial upper level $|q\rangle$ and de-excites to a lower level $|p\rangle$ with energy $E_p = 13.6\,eV/p^2$ and energy difference $E_{pq} = 13.6eV\left(1/p^2 - 1/q^2\right)$, then the de-excitation rate coefficient is given by

$$K_{qp} = \frac{1.6\times10^{-7}\left(kT_e\right)^{0.5}\, g_p/g_q}{kT_e + \Gamma_{pq}} \times \left[A_{pq}\ln\left(\frac{0.3\,kT_e}{\Re} + \Delta_{pq}\right) + B_{pq}\right], \qquad (5.)$$

where, $g_p = 2p^2$ and $g_q = 2q^2$ are the statistical weights, and defining $s=|p-q|$,

$$\Delta_{pq} = \exp\left(\frac{-B_{pq}}{A_{pq}}\right) + \frac{0.06s^2}{pq^2} \quad \text{and}$$

$$\Gamma_{pq} = \frac{\Re\,\ln\left(1+\frac{q^3\,kT_e}{\Re}\right)\left[3+11\left(\frac{s}{q}\right)^2\right]}{\left(6+1.6ps+\frac{0.3}{s^2}+\frac{0.8\,p^{1.5}}{s^{0.5}}\left|s-0.6\right|\right)},$$

$$A_{pq} = \left(\frac{2\Re}{E_{pq}}\right) f_{pq} \quad \text{and}$$

$$B_{pq} = \frac{4\Re^2}{p^3}\left(\frac{1}{E_{pq}^2} + \frac{4\,E_{pq}}{3E_{pq}^3} + b_p\,\frac{E_p^2}{E_{pq}^4}\right) \quad \text{with}$$

$$b_p = \frac{1.4\ln p - 0.7}{p} - \frac{0.51}{p^2} + \frac{1.16}{p^3} - \frac{0.55}{p^4}.$$

The value $\Re$ is the Rydberg constant and $kT_e$ is in eV, $f_{pq}$ is the absorption oscillator strength given by Johnson (1972). For comparison SH use data from Percival & Richards (1978). Again, differences can easily be a factor of two and are a basic problem (Chang, Avrett, & Loeser 1991).

Case B does not define the populations of the $n$=1 or $n$=2 levels, so collisions from these levels are not included in the comparisons made below.

### *2.4. Radiative Cascade*

This is the most difficult process to simulate, and the only one whose treatment is non-standard, because the branching ratios from each level depends on both density and temperature. Physically, captures tend to occur to higher angular momentum, *l*, levels. As electrons cascade downward they further tend to "bunch up" at the highest possible *l* values because of the $\Delta l$=1 selection rule. Distant collisions with slow-moving protons tend to distribute the electron density population according to statistical weight. As a result, the actual distribution of populations among the *l* levels will have a density dependence, and the branching ratios from a particular level will too.

The dashed line in Figure 2 shows the Pα to Hβ ratio from SH for three temperatures. The lines have a common upper level and so the intensity ratio is proportional to the 4-3/4-2 branching ratio. The lines are observationally important since both are strong and can be readily detected, and the ratio has a large dependency on the density and temperature of the gas. There are two simple limits; the well *l*-mixed, high density limit with Pα/Hβ = 0.277 (at all temperatures) corresponding to the Seaton (1959) recombination spectrum, and the low density limit with $P\alpha/H\beta = 0.339$ (at 10,000 K) where *l* mixing collisions are negligible (corresponding to the calculations presented by Pengelly 1964 and Martin, 1988).

We defined a transition probability from each excited level in terms of a constant total Case B lifetime (the sum of the Einstein A's for the upper level minus the ground state transition), and a set of branching ratios which depend on density and temperature as the solid line in Figure 2 shows. These branching ratios were obtained by fitting the SH results, for transitions from the first excited state up to $n$ =15. Above $n$ =15, the branching ratios were left at their high density values given by the Einstein A's.

Branching ratios from the upper level $n \leq 7$ were calculated using two linear interpolations in temperature, one at $\log n_e \leq 4$, and one at $n_e = n_{min}$, where $n_{min}$ is defined as the minimum density of the branching ratio. For $n = 4$, $\log n_{min} = 9$ (see figure 2). At $\log n_e \geq 11$ (for transitions from the upper level $n = 4$) the branching ratio is constant for all temperatures since the atom has gone over to the well *l*-mixed limit (see Figure 2). For intermediate densities between the three points, $\log n_e =4$, $\log n_e = n_{min}$, and $\log n_e = 11$, we linearly interpolate the branching ratio in density.

Table 2a gives the coefficients of the fits used to calculate the branching ratios up to, and including, quantum level $n$ =7. For $\log n_e \leq 4$, the branching ratio can be fit with the equation $A + B\ln T_e$, and at $\log n_e = n_{min}$ in column 3 of the table the branching ratio can be approximated with $D + E\ln T_e$. The value $C$ in column 6 in the table is the high density branching ratio, good for $\log n_e \geq n_{min} +2$. The temperature range for all fits of the branching ratio is 1,000 K to 30,000 K, the valid range of SH.

Table 2b gives the coefficients of the fits used to calculate the branching ratios for the atomic levels $n \geq 8$. For these levels the branching ratios were assumed to be simpler than Table 2a (since they are more nearly constant), such that the fits only interpolate in temperature for $\log n_e \leq 4$ ($n_{min}$ is not defined). The upper level is denoted in column (1) of the table and the transitions are listed horizontally to the right. The notation is the same as in Table 2a: at $\log n_e \leq 4$, the branching ratio is $A + B\ln T_e$, and for log $n_e \geq 11$, the branching ratio is $C$. For intermediate densities we linearly interpolate in density to calculate the branching ratio.

Errors in the branching ratios at high densities are generally less than 1%, typically less than 0.5% since the branching ratios go to the high density limit for all temperatures. Errors at low density are larger, on the order of 1%-2%. The intermediate density range, $4 \leq \log n_e \leq 9$, errors are still larger, typically 5%-10%, because of deviation from linearity of the branching ratios for highly excited levels. In what follows these errors will be small compared with other uncertainties.

## 3. Results

The atom has been incorporated into the radiative equilibrium code CLOUDY (Ferland 1996). We have left the total number of levels of the hydrogen atom as an option. More levels generally produce a better agreement with the SH results, but at the expense of longer execution times. Tests show that lower temperatures require more levels, because the highest level must be well within $kT_e$ of the continuum for three-body recombination to achieve its full efficiency. A flexible choice in the number of levels was a major reason for the chosen structure of the model atom. For all of the results shown here a 50 level atom was used.

Figure 3a shows a contour plot of the emissivity of Hβ (in units of $4\pi j_\nu/n_e n_p$) for temperatures greater than $10^3$ K and the full range of densities considered by SH. The Hβ emissivity varies by 5 orders of magnitude for this range in conditions. Figure 3b shows a ratio of this work and that of SH. For nebular temperatures (5000 K – 20,000 K) and all densities the differences are less than 2%. For low temperatures (< 3,000 K) and low densities ($<10^7$ cm$^{-3}$) the general agreement is to within 6%. In this limit the differences are due to our use of well *l*-mixed Einstein A's for the higher levels (see section 2.4). At low temperatures captures are mainly to these levels, which are not actually mixed at low densities. This is a basic limitation of our approach.

Uncertainties in the collision data (see section 2.2) are the main reason our atom does not agree exactly with SH at low temperatures (<3000 K) and high densities (>$10^8$ $cm^{-3}$). We quantify the basic uncertainties by modifying the collisional rate coefficients from Vriens & Smeets by a factor of 2, the representative uncertainty discussed by Chang, Avrett, & Loeser (1991). Figure 3c shows the ratio of our predictions with and without this scale factor, and so shows effects of these uncertainties. Results changed by nearly a factor of two at low temperatures and high densities, in response. We attribute the differences in our results and SH at low temperature and high densities to the fact that we use different sources for the collision data. The results for nebular conditions (5000 K - 20,000 K) are affected very little, and we agree with SH very well.

Figure 4 shows our results with SH for two Balmer lines at $T_e = 10^4$ K relative to Hβ. At log $n_e$ = 2.0 we calculate the Balmer decrement, Hα/Hβ/Hγ/Hδ, to be 2.866/1.0/0.473/0.262 and at log $n_e$ = 14.0, it is 3.548/1.0/0.388/0.189. The SH results for the same densities are 2.86/1.0/0.468/0.259 and 3.41/1.0/0.387/0.188, respectively.

Figure 5 shows the intensities of two observationally important infrared emission lines at $T_e = 10^4$ K relative to Hβ. They agree with SH to typically within 1-2%, with the largest errors being 5%. Tests show that this accuracy is typical for lines produced by levels lower than 15, the highest level we attempted to take account of variable *l*-mixing.

The Case B approximation does not include the effects of the ground and the first excited states by design, so these results cannot be directly applied to dense clouds. Among the processes which were disabled for the comparisons above were continuum pumping, photoionization and induced recombination from excited states, and collisional excitation from the ground or first excited states. For dense clouds such as the BLR of a quasar, these are very important processes.

Table 3 compares Case B with this work for $10^4$ K and an electron density of $10^{11}$ $cm^{-3}$, typical BLR conditions Ferland et al. (1992). Column 1 lists the line labels, column 2 are the Case B ratios (SH) with respect to Hβ, column 3 are results from our compact atom with all induced processes disabled. Columns 4 and 5 show the effects of multiplying the collision data for levels higher than 2 by a factor of 2 and 0.5 respectively. Clearly the results are sensitive to these uncertain numbers. Column 6 shows the results of enabling induced photoionization and recombination and stimulated emission for a continuum shape and an ionization parameter (ratio of photon density to hydrogen density) of 0.1, given by the standard BLR model from Ferland et al. (1992). Column 7 has collisions from the ground and first excited states included in the calculation. Column 8 lists the results of allowing the Lyman lines to be optically thin.

Finally, the results of a complete calculation are shown in column 9. This solves for the energy balance and so has a depth dependent temperature (the mean is close to $10^4$ K), it again assumes a hydrogen density of $10^{11}$ $cm^{-3}$, has a finite column density of $10^{26}$ $cm^{-2}$, and corresponding H line optical depths, and a non-thermal active galactic nuclei continuum. In the last case most hydrogen lines are optically thick, and both photoionization and collisional ionization from excited levels are very important. Clearly the line spectrum is far from Case B. This underscores the influence of the ground and first excited state in high density situations.

In summary, we find that a compact hydrogen atom can reproduce quite well the hydrogen emission spectrum calculated with a more extensive model atom at most temperatures and densities. Our results underscore the points made by Chang, Avrett, & Loeser (1991), indicating the need for more accurate collisional rate coefficients.

The authors wish to thank K.T. Korista and D.A. Verner for very insightful discussion and questions throughout this work. We also thank the referee, G. Shields, for his insightful comments on our manuscript. Research in Nebular Astrophysics at the University of Kentucky is supported by the NSF through grant AST 93 -19034, and by NASA with award NAG 5-3223.

## 5. Figures

Figure 1. Total Case B recombination coefficient defined as the radiative recombination plus three-body recombination minus collisional ionization compared with SH. This work is shown as solid lines and SH as dashed lines. Three temperatures are shown, 5,000 K, 10,000 K, and 20,000 K. The differences are due to different sources of collisional ionization rate coefficients.

Figure 2. The SH (dashed lines) predicted Pα to Hβ ratio as a function of density and temperature. Three temperatures are shown, 5,000 K, 10,000 K, and 20,000 K moving from top to bottom. Our fit to the branching ratio is shown as a solid line.

Figure 3. Contour plots. a) shows the log of the total Hβ emissivity (in units of $4\pi j_{\nu}/n_e n_p$) for all densities and temperatures. Solid lines are 1 dex increments and dashed lines are 0.2 dex steps. b) is the ratio of our predicted Hβ emissivity to SH, 10% increments are solid lines and 2% steps are dashed lines. Part c) of the figure is the same as b), but with the collisional atomic data multiplied by 2, as described in the text. In part c) the solid lines are 20% differences and the dashed 10%. Factors of two changes result for some parameters.

Figure 4. Selected ratios of Balmer lines relative to Hβ for $10^4$ K and the full range of density. This work is shown as solid lines and the SH results as dashed lines.

Figure 5. Two examples of important infrared lines shown at $10^4$ K and the full range of density. This work is shown as solid lines and the SH results as dashed lines. Shown are a) Br γ to Hβ, b) Hu α to Hβ. Hu α is Humphries α the transition from level 7 to 6 in hydrogen.

## 6. Tables

TABLE 1

| | 1 | 2 | 3 | 4 | 5 | 6 | 7 | 8 |
|---|---|---|---|---|---|---|---|---|
| a | -10.78145 | -11.04340 | -11.20313 | -11.31452 | -11.40622 | -11.48461 | -11.46468 | -11.50326 |
| b | -0.38890 | -0.39351 | -0.42458 | -0.43939 | -0.43647 | -0.44246 | 0.03271 | -0.17837 |
| c | 4.68469 | 4.84334 | 5.25370 | 5.45398 | 5.46028 | 5.56951 | -0.07602 | 2.30302 |
| d | 0.06404 | 0.06921 | 0.08192 | 0.08538 | 0.08639 | 0.08998 | 0.00660 | 0.05393 |
| e | -0.87423 | -0.95686 | -1.12613 | -1.16660 | -1.18431 | -1.24103 | 0.02975 | -0.53187 |
| f | -0.00510 | -0.00550 | -0.00682 | -0.00740 | -0.00737 | -0.00755 | 0.01032 | -0.00120 |
| g | 0.08141 | 0.09216 | 0.11422 | 0.11806 | 0.11914 | 0.12544 | -0.10863 | 0.0 |
| h | 0.00248 | 0.00306 | 0.00412 | 0.00404 | 0.00412 | 0.00449 | -0.00253 | 0.0 |
| i | -0.03877 | -0.05025 | 0.0 | -0.06488 | -0.06752 | -0.07536 | 0.0 | 0.0 |

| | 9 | 10 | 11 | 12 | 13 | 14 | 15 | Case B |
|---|---|---|---|---|---|---|---|---|
| a | -11.71000 | -11.61098 | -11.99276 | -11.71513 | -11.66252 | -11.69889 | -11.71537 | -9.97652 |
| b | 8.07670 | -0.16697 | 4.76460 | -0.14672 | 0.02722 | 0.02892 | 0.02312 | 0.03506 |
| c | -92.51728 | 2.20472 | -54.29910 | 2.02504 | -0.28688 | -0.28812 | -0.28765 | 0.15861 |
| d | -1.46915 | 0.05915 | -0.88171 | 0.06253 | 0.09368 | 0.09033 | 0.11802 | -0.03762 |
| e | 18.56208 | -0.59021 | 10.94495 | -0.63483 | -0.71041 | -0.69566 | -0.94380 | 0.30113 |
| f | 0.51035 | -0.00135 | 0.33187 | -0.00146 | 0.00046 | 0.00173 | -0.00225 | 0.00762 |
| g | -4.93019 | 0.0 | -3.22476 | 0.0 | -0.05483 | -0.06497 | -0.03708 | -0.06397 |
| h | -0.12090 | 0.0 | -0.07931 | 0.0 | -0.00014 | -0.00048 | 0.00047 | -0.00023 |
| i | 0.0 | 0.0 | 0.0 | 0.0 | 0.0 | 0.0 | 0.0 | 0.00127 |

TABLE 2a

Fits of the branching ratios.

| q | p | $n_{min}$ | A | B | C | D | E |
|---|---|---|---|---|---|---|---|
| (1) | (2) | (3) | (4) | (5) | (6) | (7) | (8) |
| 4 | 2 | 9 | -0.1211 | 0.0603 | 0.4835 | 0.3618 | 0.0146 |
| | 3 | | 1.1216 | -0.0604 | 0.5165 | 0.6382 | -0.0146 |
| 5 | 2 | 8 | -0.2310 | 0.0555 | 0.3407 | 0.1284 | 0.0220 |
| | 3 | | 0.1541 | 0.0136 | 0.2961 | 0.2895 | 0.0000 |
| | 4 | | 1.0777 | -0.0692 | 0.3632 | 0.5808 | -0.0218 |
| 6 | 2 | 8 | -0.2393 | 0.0490 | 0.2740 | 0.1240 | 0.0152 |
| | 3 | | -0.0180 | 0.0226 | 0.2190 | 0.1822 | 0.0033 |
| | 4 | | 0.2895 | -0.0061 | 0.2180 | 0.2562 | -0.0041 |
| | 5 | | 0.9860 | -0.0668 | 0.2900 | 0.4378 | -0.0145 |
| 7 | 2 | 7 | -0.2313 | 0.0441 | 0.2360 | -0.0102 | 0.0238 |
| | 3 | | -0.0752 | 0.0244 | 0.1800 | 0.0889 | 0.0083 |
| | 4 | | 0.1399 | 0.0024 | 0.1636 | 0.1619 | 0.0000 |
| | 5 | | 0.3215 | -0.0135 | 0.1750 | 0.2684 | -0.0090 |
| | 6 | | 0.8889 | -0.0615 | 0.2450 | 0.4915 | -0.0232 |

TABLE 2b

Fits of the branching ratios for upper levels greater than 7

| | 2 | 3 | 4 | 5 | 6 | 7 | 8 | 9 | 10 | 11 | 12 | 13 | 14 |
|---|---|---|---|---|---|---|---|---|---|---|---|---|---|
| 8 | | | | | | | | | | | | | |
| A | -0.2811 | -0.0850 | 0.0523 | 0.1861 | 0.3673 | 0.7604 | | | | | | | |
| B | 0.0465 | 0.0227 | 0.0080 | -0.0049 | -0.0206 | -0.0518 | | | | | | | |
| C | 0.2096 | 0.1565 | 0.1352 | 0.1322 | 0.1490 | 0.2175 | | | | | | | |
| 9 | | | | | | | | | | | | | |
| A | -0.2699 | -0.0942 | 0.0149 | 0.1092 | 0.2100 | 0.3574 | 0.6726 | | | | | | |
| B | 0.0436 | 0.0220 | 0.0098 | 0.0002 | -0.0091 | -0.0214 | -0.0452 | | | | | | |
| C | 0.1923 | 0.1409 | 0.1181 | 0.1095 | 0.1120 | 0.1310 | 0.1961 | | | | | | |
| 10 | | | | | | | | | | | | | |
| A | -0.2548 | -0.1023 | -0.0077 | 0.0668 | 0.1332 | 0.2143 | 0.3400 | 0.6768 | | | | | |
| B | 0.0406 | 0.0216 | 0.0108 | 0.0029 | -0.0036 | -0.0108 | -0.0209 | -0.0413 | | | | | |
| C | 0.1783 | 0.1293 | 0.1063 | 0.0955 | 0.0929 | 0.0984 | 0.1185 | 0.1807 | | | | | |
| 11 | | | | | | | | | | | | | |
| A | -0.2481 | -0.1040 | -0.0170 | 0.0415 | 0.0840 | 0.1445 | 0.2123 | 0.3157 | 0.5575 | | | | |
| B | 0.0390 | 0.0209 | 0.0108 | 0.0043 | 0.0000 | -0.0058 | -0.0115 | -0.0195 | -0.0368 | | | | |
| C | 0.1683 | 0.1206 | 0.0980 | 0.0861 | 0.0807 | 0.0812 | 0.0885 | 0.1082 | 0.1684 | | | | |
| 12 | | | | | | | | | | | | | |
| A | -0.2370 | -0.0988 | -0.0240 | 0.0272 | 0.0740 | 0.1039 | 0.1508 | 0.2043 | 0.2924 | 0.5170 | | | |
| B | 0.0370 | 0.0197 | 0.0108 | 0.0051 | 0.0000 | -0.0029 | -0.0072 | -0.0445 | -0.0180 | -0.0337 | | | |
| C | 0.1598 | 0.1138 | 0.0915 | 0.0791 | 0.0727 | 0.0685 | 0.0750 | 0.0802 | 0.1004 | 0.1586 | | | |
| 13 | | | | | | | | | | | | | |
| A | -0.2485 | -0.1017 | -0.0311 | 0.0142 | 0.0660 | 0.0680 | 0.1112 | 0.1518 | 0.1983 | 0.2732 | 0.4787 | | |
| B | 0.0375 | 0.0195 | 0.0110 | 0.0058 | 0.0000 | 0.0000 | -0.0043 | -0.0079 | -0.0115 | -0.0168 | -0.0310 | | |
| C | 0.1528 | 0.1082 | 0.0863 | 0.0738 | 0.0669 | 0.0632 | 0.0631 | 0.0662 | 0.0746 | 0.0941 | 0.1505 | | |
| 14 | | | | | | | | | | | | | |
| A | -0.2237 | -0.0995 | -0.0400 | 0.0019 | 0.0600 | 0.0610 | 0.0640 | 0.1128 | 0.1464 | 0.1811 | 0.2579 | 0.4421 | |
| B | 0.0344 | 0.0188 | 0.0115 | 0.0065 | 0.0000 | 0.0000 | 0.0000 | -0.0051 | -0.0079 | -0.0103 | -0.0159 | -0.0281 | |
| C | 0.1468 | 0.1037 | 0.0820 | 0.0697 | 0.0623 | 0.0581 | 0.0564 | 0.0572 | 0.0611 | 0.0698 | 0.0890 | 0.1437 | |
| 15 | | | | | | | | | | | | | |
| A | -0.2372 | -0.0955 | -0.0370 | -0.0011 | 0.0560 | 0.0560 | 0.0570 | 0.0979 | 0.1152 | 0.1354 | 0.18100 | 0.2459 | 0.4315 |
| B | 0.0354 | 0.0183 | 0.0108 | 0.0065 | 0.0000 | 0.0000 | 0.0000 | -0.0043 | -0.0058 | -0.0072 | -0.0108 | -0.0152 | -0.0260 |
| C | 0.1420 | 0.1000 | 0.0790 | 0.0660 | 0.0585 | 0.0540 | 0.0520 | 0.0510 | 0.0530 | 0.0566 | 0.0655 | 0.0850 | 0.1380 |

Table 3

| line | Case B[a] | 1[b] | 2[c] | 3[d] | 4[e] | 5[f] | 6[g] | BLR[h] |
|---|---|---|---|---|---|---|---|---|
| (1) | (2) | (3) | (4) | (5) | (6) | (7) | (8) | (9) |
| Hβ[i] | 1.84(-25) | 1.80(-25) | 1.95(-25) | 1.64(-25) | 2.94(-25) | 3.03(-25) | 1.25(-25) | 1.04(-27) |
| Lα | 31.5 | -- | -- | -- | -- | -- | 34.68 | 65.6 |
| Hα | 2.55 | 2.56 | 2.51 | 2.61 | 6.71 | 7.11 | 2.19 | 3.28 |
| Hγ | 0.570 | 0.570 | 0.594 | 0.545 | 0.4511 | 0.443 | 0.630 | 0.392 |
| Hδ | 0.394 | 0.395 | 0.406 | 0.363 | 0.268 | 0.262 | 0.480 | 0.211 |
| H10 | 0.0959 | 0.0985 | 0.0906 | 0.101 | 0.0596 | 0.0578 | 0.138 | 0.104 |
| Pα | 0.277 | 0.277 | 0.277 | 0.277 | 0.277 | 0.277 | 0.277 | 0.205 |
| Pβ | 0.168 | 0.169 | 0.175 | 0.160 | 0.133 | 0.130 | 0.185 | 0.140 |
| Brα | 0.0654 | 0.0651 | 0.0679 | 0.0623 | 0.0515 | 0.0506 | 0.0719 | 0.0301 |
| Brγ | 0.0351 | 0.0356 | 0.0347 | 0.0336 | 0.0220 | 0.0214 | 0.0465 | 0.0301 |
| Huα | 0.0092 | 0.0093 | 0.0091 | 0.0088 | 0.0058 | 0.0056 | 0.0122 | 0.0022 |

[a] Case B results from SH

[b] Cloudy with physical procesess disabled as described in text.

[c] Same as column 3, but with collision data times 2.

[d] Same as column 3, but with collision data times 0.5.

[e] induced processes enabled.

[f] Collisions between 2s and 2p electrons included.

[g] Case B assumption turned off.

[h] Simple BLR cloud as described in the text

[i] Hβ emissivity in units of erg $cm^3$ $sec^{-1}$, number in paratheses is the exponent.